\magnification=\magstep2
\centerline {\bf Generating  Bounds for the Ground State Energy of the}
\centerline {\bf Infinite Quantum Lens Potential  }
\bigskip
\centerline {\bf Carlos R. Handy}
\centerline {Department of Physics and}
\centerline {Center for Theoretical Studies of Physical Systems}
\centerline {Clark Atlanta University}
\centerline {Atlanta, Georgia 30314}

\bigskip
\centerline {\bf C. Trallero-Giner and Arezky H. Rodriguez}
\centerline { Department of Theoretical Physics}
\centerline {University of Havana}
\centerline {10400 Havana, Cuba}
\baselineskip 30 true pt
\vfil\eject
%*******DEFINITIONS FOR TABLE COMMANDS *************************
\newdimen\tempdim
\newdimen\othick   \othick=.4pt
\newdimen\ithick    \ithick=.4pt
\newdimen\spacing   \spacing=9pt
\newdimen\abovehr   \abovehr=6pt
\newdimen\belowhr   \belowhr=8pt
\newdimen\nexttovr  \nexttovr=8pt

\def\rr{\hfil\down{\abovehr}&\omit\vrsp\vrule width\othick\cr
\noalign{\hrule height\ithick}\up{\belowhr}&}
\def\up#1{\tempdim=#1\advance\tempdim by1ex
\vrule height\tempdim width0pt depth0pt}
\def\down#1{\vrule height0pt depth#1 width0pt}
\def\large#1#2{\setbox0=\vtop{\hsize#1 \lineskiplimit=0pt \lineskip=1pt
\baselineskip\spacing \advance\baselineskip by 3pt \noindent
        #2}\tempdim=\dp0\advance\tempdim by\abovehr\box0\down{\tempdim}}

\def\vrsp{\hskip\nexttovr\relax}
\def\toprule#1{\def\startrule{\hrule height#1\relax}}
\toprule{\othick}
\def\nstrut{\vrule height\spacing depth3.5pt width0pt}

\def\preamble#1{\def\startup{#1}}
\preamble{&##}
{\catcode`\!=\active
\gdef!{\hfil\vrule width0pt\vrsp\vrule width\ithick\relax\vrsp&}}

\def\table #1{\vbox\bgroup \setbox0=\hbox{#1}
        \vbox\bgroup\offinterlineskip \catcode`\!=\active
        \halign\bgroup##\vrule width\othick\vrsp&\span\startup\nstrut\cr
        \noalign{\medskip}
        \noalign{\startrule}\up{\belowhr}&}

\def\caption #1{\down{\abovehr}&\omit\vrsp\vrule width\othick\cr
        \noalign{\hrule height\othick}\egroup\egroup \setbox1=\lastbox
        \tempdim=\wd1 \hbox to\tempdim{\hfil \box0 \hfil} \box1 \smallskip
        \hbox to\tempdim{\advance\tempdim by-20pt\hfil\vbox{\hsize\tempdim
        \noindent #1}\hfil}\egroup}
%*********END OF DEFINITIONS*********************************
\bigskip\centerline {\bf Abstract}
Moment based methods have produced efficient multiscale
quantization algorithms for  solving 
singular perturbation/strong coupling 
problems. One of these, the Eigenvalue Moment Method (EMM),
 developed by Handy et al (Phys. Rev. Lett.{\bf 55}, 931 (1985);
ibid, {\bf 60}, 253 (1988b)), generates converging
 lower and upper bounds  to a specific
discrete state energy, once the signature property of the
 associated wavefunction is known. This method is particularly
effective for multidimensional, bosonic ground state problems,
 since the corresponding wavefunction must be of uniform signature,
 and can be taken to be positive.   Despite 
this, the vast majority of problems studied have been
 on unbounded domains. 
The important problem of an electron in an infinite
 quantum lens potential defines a
challenging extension of EMM to systems defined on a compact domain. 
We investigate this here, and introduce novel modifications
 to the conventional
EMM formalism that facilitate its adaptability to the required boundary
conditions.
\vfil\break
\bigskip\centerline {\bf I. Introduction}
\bigskip\indent Self-assembled quantum dots (QDs),
 obtained by interrupted growth
in strained semiconductors, offer an attractive and fascinating array of physical properties (Leonard et al (1993,1994)). Differential capacitance (Drexler et al (1994), Miller et al (1997)), magnetic-conductance (Medeiros-Ribeiro et al (1997)), and optical experiments (Fafard et al (1994), Lee et al (2000)) demonstrate that electronic states are strongly confined inside such structures.

 Typically, a lens 
geometry is assumed (Leonard et al (1994)), with a circular cross section
of maximum radius $a$, and maximum thickness $b$; wherein, the charge carriers are
confined by a  hard
wall (infinite) potential.
The mathematical 
characterization of the energy levels of such nanostructures  is a delicate
problem, particularly in the thin lens limit ${b\over a} \rightarrow 0$, 
which corresponds to a singular perturbation regime.

\bigskip\indent Recently, conformal analysis methods were used to 
solve the infinite quantum
 lens potential (Rodriguez et al (2001)).
 Preliminary results underscore the delicate nature of the 
thin lens regime. In order to better assess the accuracy of such methods,
we have developed
 an eigenenergy bounding procedure that, at low order,
yields exceptionally tight bounds to the discrete state energy levels.  The 
details are presented here, with respect to the ground state.

Our bounding procedure is based on the 
Eigenvalue Moment Method (EMM) formalism of Handy et al (1985,1988a,b). This,
linear programming based (Chvatal (1983)),
formalism has been shown to be exceptionally well
suited for singular perturbation/strong coupling problems. It is very simple to
use, and involves the application of
 fundamental theorems arising from the classic {\it Moment Problem} (Shohat and Tamarkin (1963), Akhiezer (1965)), as well as 
theorems pertaining to the signature structure of bosonic (ground state)
 wavefunctions (Reed and Simon (1978)). 

Specifically,  the multidimensional bosonic ground state wavefunction 
must be  of uniform signature, which can be taken to be positive:
$$ \Psi_{gr}({\overrightarrow r}) > 0 . \eqno (1)$$
It will then satisfy the positive integral relations:

$$ \int\int\int \ dx dy dz \Big( {\cal P}_{C}({\overrightarrow r})\Big)^2
\Psi_{gr}({\overrightarrow r}) > 0, \eqno(2)$$
where ${\cal P}_{C} \equiv \sum_{l,m,n}C_{l,m,n}x^ly^mz^n$ is an arbitrary
 polynomial. In terms of the power moments, $\mu(p_1,p_2,p_3) 
\equiv \int\int\int \ dx dy dz\ x^{p_1}y^{p_2}z^{p_3}\ \Psi_{gr}
({\overrightarrow r})$, for nonnegative integer $p_i$'s,
 these integrals become the Hankel-Hadamard (HH),
quadratic form,
inequalities:

$$ \sum_{l_1,m_1,n_1}\sum_{l_2,m_2,n_2} 
C_{l_1,m_1,n_1 } \mu(l_1+l_2,m_1+m_2 ,n_1+n_2)
 C_{l_2,m_2,n_2 } > 0, \eqno(3)$$
for arbitrary $C$'s (not all identically zero).

The Fourier transform of the Schrodinger equation  usually
 admits a power series
expansion, whose coefficients (i.e. the moments) satisfy a linear 
recursion relation, referred to as the Moment Equation (ME). This relation
exists for any energy parameter value, $E$.
The entire set of power moments is divided into two subsets:
$$\{\mu({\overrightarrow p})| \ \forall {\overrightarrow p}\} =
 \{\mu({\overrightarrow \ell})| 
{\overrightarrow \ell} \in {\cal M}_s\} \bigcup \{\mu({\overrightarrow p})|
{\overrightarrow p} \notin {\cal M}_s\}.\eqno(4)$$
The first subset corresponds to
the initialization moments, or {\it missing moments}, which must be specified 
before all of the other moments can be generated, through the ME relation.

The generated moments, those
in the second subset, are linearly dependent on the missing moments.
We can  represent the ME relationship as
$$ \mu({\overrightarrow {p}}) = \sum_{{\overrightarrow {\ell}} \in {\cal M}_s}
M_E({\overrightarrow{p}},{\overrightarrow {\ell}}) 
\mu({\overrightarrow {\ell}}), \eqno(5)$$
${\overrightarrow p} \notin {\cal M}_s$.
The $M_E$ coefficients are dependent on $E$, and can be defined
so that the above is also
 valid for the missing moments
as well (i.e. $M_E({\overrightarrow {\ell_1}},{\overrightarrow {\ell_2}}) =
\delta_{{\overrightarrow {\ell_1}},{\overrightarrow {\ell_2}}  }$,
for ${\overrightarrow \ell}_{1,2} \in {\cal M}_s$).

For one dimensional systems, the number of missing moments is finite, and
denoted $1+m_s$. For multidimensional systems, the number of missing moments
is infinite; however, at any point in the calculation, one works with a finite
number of them, which in turn determine a finite number of the generated 
moments. Generally, there will be many times more generated moments, than the
corresponding number of missing moments.

Since the Moment Equation is a homogeneous relation, one must impose a normalization condition.
This is normally done with respect to the missing moments. For instance,
we can take
$$\sum_{\overrightarrow {\ell}} \mu({\overrightarrow \ell}) = 1.\eqno(6)$$
 Assuming 
this, one can substitute the ME relation, for the generated moments,
 into the HH inequalities. Since all the moments
are linear in the missing moments, a linear programming problem is
defined of the form
$$ \sum_{{\overrightarrow {\ell}}} {\Lambda}_{{\overrightarrow {\ell}}}(E;C)
\mu({{\overrightarrow {\ell}}}) > 0, \eqno(7)$$
\smallskip\noindent where the ${\Lambda}$ coefficients are nonlinearly dependent
on $E$, and quadratically dependent on the (arbitrary) $C$'s.

In practice, we work within a finite dimension ($I$) subspace for the $C$
coefficients. Define this by  
$(l,m,n) \in {\cal C}^{(I)}$. The required moments are
$\{\mu({\overrightarrow p})|
{\overrightarrow p} = (l_1,m_1,n_1)+(l_2,m_2,n_2)$, where
 $(l_{1,2},m_{1,2},n_{1,2})
\in{\cal C}^{(I)}\}$. One must then determine the missing moments that generate
these. They in turn define the linear programming variable space in Eq.(7).

For a given dimension, $I$, at an arbitrary energy value, $E$,
the HH inequalities 
will either have a missing moment solution set, ${\cal U}_E^{(I)}$,
or not ${\cal U}_E^{(I)} = \oslash$. If there is a solution set,
it must be convex. This convex set may be considered as the intersection of
an (uncountably) infinite number of polytopes (convex sets formed from the
intersection of a finite number of hyperplanes).

The objective of the linear programming based,
 algorithmic implementation of EMM, is to  quickly determine the existence or 
nonexistence of ${\cal U}_E^{(I)}$. At any order $I$, the feasible energy
values (those for which the convex set exists)
 define an energy interval, $(E_L^{(I)},E_U^{(I)})$, within which the 
true ground state value, $E_{gr}$, must lie. As the order is increased, 
the energy endpoints define the converging lower and upper bounds to the
physical answer:

$$  E^{(I)}_{L} < E^{(I+1)}_{L} < \ldots < E_{gr} < \dots < E^{(I+1)}_{U} < E^{(I)}_{U} 
,\eqno(8)$$
\smallskip\noindent $I \rightarrow \infty$.

The EMM formalism was used to generate rapidly
converging bounds to the ground state binding energy for
hydrogenic atoms in superstrong magnetic fields (Handy et al (1988a,b)),
otherwize known  as the {\it Quadratic Zeeman effect}.
This problem had been notoriously difficult, yielding varying
results depending on the method used. The ability of EMM to define
tight bounds  to the ground state binding energy enabled one to discriminate
between competing (energy estimation) methods. In particular, it confirmed
the results of LeGuillou and Zinn-Justin (1983),
which were based on order dependent, conformal analysis.

The consistency of the 
EMM generated results, and those based on conformal analysis,
is more than a coincidence. In one dimension, an affine map transformation
of the point $x$
is defined by $x \rightarrow {{x-{\tau}}\over {s}}$, where $s$ and $\tau$
are scale and translation parameters, respectively. An affine map transform
of a given function, ${\cal P}(x) \rightarrow {\cal P}({{x-{\tau}}\over {s}})$,
corresponds to a translation and stretching (or contraction).

The  
variational procedure inherent to EMM,
is, in fact, affine invariant. This is immediately clear from Eq.(2), since the
variation samples over all polynomial functions, and the
 space of polynomials
is invariant under affine transforms.
To this extent, EMM is in keeping with the underlying philosophy of conformal
analysis, and should yield either consistent, or better,
results.

This affine map invariance underscores the fundamental
complementarity between
Moment Quantization  methods, such as EMM, and explicitly multiscale methods
such as Wavelet Transform theory (Handy and Murenzi (1998)).  This further
confirms the relevancy of EMM to singular perturbation type problems which
require a careful balancing of large and small scale contributions.

Despite the numerous types of problems the  EMM formalism has been applied
to, as reviewed in the
cited references, it has not been used on problems defined on a compact domain.
Such problems require a modification of the basic formalism, in order to
adapt them to the required boundary conditions. The infinite quantum lens
potential is one such important case. 

Before applying the EMM analysis to this problem, we
discuss certain preliminary, pedagogic, examples, in order to facilitate
the more comprehensive analysis that follows.
We provide a short review of EMM
 by
considering two, relatively simple, problems. The first is the sextic anharmonic
 oscillator
problem, in one dimension. The second is the infinite square well potential.
These will introduce us to the necessary linear programming, HH, relations for
Hamburger, Stieltjes, and Hausdorff moment problems. The infinite square well
problem is an example of the latter, which also corresponds to the infinite lens potential. We also discuss how to elliminate possible boundary term contributions to the Moment Equation, and still be consistent with EMM theory. This analysis has not been presented elsewhere, and defines an important, new contribution,
with respect to the EMM formalism.
\vfil\break
\bigskip\centerline {\bf II. The Sextic Anharmonic Oscillator}
\bigskip\noindent{\underbar{\it Hamburger Moment Formulation}}
\bigskip\indent Consider the sextic anharmonic oscillator potential problem:

$$-\epsilon \partial_x^2\Psi(x) + \Big ( mx^2 + g x^6 \Big ) \Psi(x) = E \Psi(x)
,\eqno(9)$$
\smallskip\noindent where the kinetic energy perturbation parameter, $\epsilon$,
 is
explicitly noted, for later reference. The mass and coupling strength 
parameters are denoted by $m$ and $g$, respectively.

\smallskip\indent The signature structure for the ground ($\Psi_0$) and first 
excited ($\Psi_1$)
 states are
known a  priori: $\Psi_i(x) = x^i\Upsilon_i(x)$,  where $\Upsilon_i(x) > 0$. For
 simplicity, we
confine our analysis to the ground state case.

\smallskip\indent Define the Hamburger power moments
$$\mu(p) = \int_{-\infty}^{+\infty}dx \ x^p \Psi(x),\eqno(10)$$
\smallskip\noindent $p \geq 0$. Upon multiplying both sides of the Schrodinger 
equation
by $x^p$, and performing the necessary integration by parts, we obtain the 
Moment Equation (ME)

$$g\mu(p+6) = -m\mu(p+2) + E \mu(p) + \epsilon\ p(p-1)\mu(p-2),\eqno(11)$$
\smallskip\noindent for $p \geq 0$. This corresponds to an effective sixth order
 finite
difference equation, in which specification of the ``initialization" moments,
 or 
{\it missing moments},
$\{\mu(\ell)|0 \leq \ell \leq 5\}$, as well as the energy parameter, $E$,
generates all of the remaining moments.
\smallskip\indent One important aspect about working within a moments' 
representation
is that
kinetic energy expansions become regular (i.e. $\epsilon$ expansions). This is 
not the case
in configuration space, requiring the use of singular perturbation type methods (Bender and Orszag (1978)).
 One immediate
impact of the regularity in $\epsilon$ is that the order of the moment equation
does not change
for $\epsilon = 0$ and $\epsilon = 0^+$ (unlike in configuration space, where 
the order of
the differential equation abruptly changes from 0 to 2).
\smallskip\indent Let us denote the missing moment order by $1+m_s$,
 where $m_s = 5$. We can express the linear dependence of the
moments on the missing moments through the expression
$$\mu(p) = \sum_{\ell = 0}^{m_s }M_E(p,\ell)\mu(\ell),\eqno(12)$$
\smallskip\noindent where
$$M_E(\ell_1,\ell_2) = \delta_{\ell_1,\ell_2}, \eqno(13)$$
for $0 \leq \ell_1,\ell_2 \leq m_s$. The $M_E$ coefficients are readily 
obtainable, since they satisfy the ME relation with respect to the $p$-index, 
in addition to the preceeding initialization condtions.
\smallskip\indent We must also impose some, convenient, normalization condition.
 This can be chosen
to be
$$ \sum_{\ell = 0}^{m_s}\mu(\ell) = 1.\eqno(14)$$
\smallskip\noindent Constraining the zeroth order
 moment, $\mu(0) = 1-\sum_{\ell = 1}^{m_s}\mu(\ell)$,we redefine the 
moment-missing moment relation as
$$\mu(p) = \sum_{\ell = 0}^{m_s} {\hat M}_E(p,\ell){\hat \mu}(\ell), \eqno (15)
$$
\smallskip\noindent where
$$ {\hat \mu}(\ell) = \cases { 1, \ for \ \ell = 0, \cr
\mu(\ell), \ for \ 1 \leq \ell \leq m_s \cr }, \eqno(16)$$ 
\smallskip\noindent and
$${\hat M}_E(p,\ell) = \cases { M_E(p,0), \ for \ \ell = 0, \cr
                                M_E(p,\ell)-M_E(p,0), \ for \ 1\leq \ell \leq m_
s \cr}. \eqno (17)$$

\smallskip\indent From the Moment Problem, we know that the moments of a 
nonnegative measure,
on the entire real axis,
must satisfy the Hankel-Hadamard (HH) constraints

$$\int_{-\infty}^{+\infty} dx \ \Big(\sum_{i=0}^IC_ix^i \Big )^2 \Psi(x)  \geq 0
, \eqno(18)$$
\smallskip\noindent for arbitrary $C_i$'s (not all zero), and $ 0 \leq I < 
\infty$.
The
zero equality is only possible for configurations made up of a finite number of
Dirac
distributions.

\smallskip\indent The HH integral constraints can be transformed into the 
quadratic 
form expression

$$ \sum_{i_1,i_2 = 0}^I C_{i_1} \mu(i_1+i_2) C_{i_2} > 0.\eqno(19)$$

\smallskip\noindent These inequalities do not guarantee uniqueness for $\Psi$ (i
.e. that
the physical solution is the only one with these moments);
 however, because
we are implicitly working with the moments of a physical system, for which there
 is uniqueness,
the nature of the ${\hat M}_E(p,\ell)$
matrix coefficients should guarantee uniqueness as well, within the moments'
representation, i.e. satisfaction of the Carlemann conditions, etc (for 
further details see Bender and Orszag (1978)).

One can then substitute the moment-missing moment relation

$$ \sum_{\ell=0}^{m_s} \Big ( \sum_{i_1,i_2 = 0}^I C_{i_1}  {\hat M}_E(i_1+i_2,
\ell)
C_{i_2} \Big ) {\hat \mu}(\ell) > 0, \eqno(20)$$
\smallskip\noindent which generates an uncountable number of linear inequalities
 (i.e. one
 linear inequality for
each $C$-tuple) in the
(unconstrained) missing moment variable space:

$$\sum_{\ell = 1}^{m_s} {\cal A}_\ell[C] \mu(\ell) < {\cal B}[C], \eqno(21)$$
\smallskip\noindent where
$${\cal A}_\ell[C] \equiv -\sum_{i_1,i_2 = 0}^I  C_{i_1}{\hat M}_E(i_1+i_2,\ell)
C_{i_2},\eqno(22)$$
\smallskip\noindent and
$${\cal B}[C] \equiv \sum_{i_1,i_2 = 0}^IC_{i_1}{\hat M}_E(i_1+i_2,0)
C_{i_2}.\eqno(23)$$
\smallskip\noindent
\smallskip\noindent
 We recall that the missing moments are restricted to $\sum_{\ell = 0}^{m_s} \mu
(\ell) = 1$.

\smallskip\indent Let ${\cal U}_{E}^{(I)}$ denote the (convex) solution set to 
the
above set of HH inequalities,
for given $E$ and $I$.
 The objective is to determine the feasible energy  interval, to order $I$,
 for which
 convex solution sets exists:
$$E \in (E_{L}^{(I)},E_{+}^{(I)}) \ {\it if} \ {\cal U}_{E}^{(I)} \neq \oslash.
\eqno(24)$$
\smallskip\noindent This can be done
through a linear programming based
 {\it cutting method} that finds the optimal $C$'s leading to a quick assesment
on the existence or non-existence of ${\cal U}_{E}^{(I)}$ (Handy et al (1988a,b)).

The preceeding formalism is appropriate if the Schrodinger equation potential 
is not symmetric. In the present case, since the potential is symmetric, we
can  define a more efficient representation by working in terms of a Stieltjes
moment formulation. This is done in the following section.

\bigskip\noindent {\underbar{\it  Stieltjes 
Moment Formulation}}
\smallskip\indent The parity invariant nature of the
 sextic anharmonic oscillator requires that the ground state be symmetric,
$\Psi(-x) = \Psi(x)$. This in turn introduces more moment constraints.

For symmetric configurations, the odd order Hamburger moments are zero,
$\mu(odd) = 0$. The even order Hamburger moments can be regarded as the moments
of a
Stieltjes measure restricted to the nonnegative real axis (through a change of 
variables,
$y = x^2$)

$$u(\rho) \equiv \mu(2\rho), \eqno(25)$$
\smallskip\noindent  where
$$u(\rho) = \int_{0}^\infty dy \ y^\rho \Phi(y), \eqno(26)$$
\smallskip\noindent and
$$\Phi(y) = {{ \Psi({\sqrt y})}\over {\sqrt y}}.\eqno(27)$$

\smallskip\indent The Stieltjes moments also satisfy a moment equation 
($\epsilon = 1$):
$$ g u(\rho+3) = -m u(\rho+1) + E u(\rho) + 2\rho(2\rho-1)\ u(\rho-1),
\eqno(28)$$
\smallskip\noindent $\rho \geq 0$.
\smallskip\indent The order of this finite difference moment equation is $1+m_s
= 3$,
leading to the representation
$$ u(\rho) = \sum_{\ell = 0}^{m_s}{\hat{\cal M}}_E(\rho,\ell) {\hat u}(\ell), 
\eqno(29)$$
where the ${\hat u}(\ell)$ are defined as before, and satisfy the constraint,
$\sum_{\ell = 0}^{m_s} u(\ell) = 1$.
\smallskip\indent One important aspect of working with Stieltjes
moments is that because the underlying function must be positive, all the
 Stieltjes
moments
must also be positive (which is not the case for the Hamburger moments). Thus,
for the adopted
normalization condition, we have

$$ 0 < u(\ell) < 1,\eqno(30)$$
\smallskip\noindent for $0 \leq \ell \leq m_s (= 2)$.

\smallskip\indent Since one is working on the nonnegative real axis, $ y \geq 0$
,
more HH constraints are possible. The constraints in Eq.(19), arising
from the integral expression in Eq.(18), define the
necessary and sufficient conditions for the moments to correspond to a 
nonnegative
measure on the entire real axis. If we pretend 
that $\Phi(y)$  exists on the entire real axis, but we want it to be zero on the
negative real axis,
 then one must also introduce the counterpart to Eq.(18) for the configuration
$y\Phi(y)$:
$$\int dy \  y^\sigma\Big(\sum_{i=0}^IC_i y^i\Big )^2 \Phi(y) > 0, \eqno(31)$$
\smallskip\noindent for $\sigma = 0,1$, and $I < \infty$.
Thus, the only way both $\Phi(y)$ and $y\Phi(y)$ can be
 nonnegative
 on the
entire $y$-axis is for $\Phi(y) = 0$, for $y < 0$. This is an intuitive way of
motivating the HH-Stieltjes moment conditions for a nonnegative measure defined
on
the nonnegative real axis. Consequently, in terms of a quadratic 
form expression, we have

$$\sum_{i_1,i_2=0}^IC_{i_1}u(\sigma+i_1+i_2) C_{i_2} > 0, \eqno(32)$$
\smallskip\noindent for $\sigma = 0,1$ and $I \geq 0$.
\smallskip\indent Repeating the same analysis presented earlier (i.e. 
substituting
 the
moment-missing moment relations, and implementing the linear programming based 
cutting
procedure) allows us to generate very tight bounds for the ground state energy.
In particular, for $\epsilon = m = g = 1$, one obtains
$$1.4356246190092 < E < 1.4356246190178, \eqno(33)$$
 for $I = 15$.
\vfil\break
\bigskip\centerline {\bf III. Defining Quantizable EMM-Moment Equations}
\smallskip\indent We now focus on issues of relevance to the application of EMM
to the
quantum lens problem. Consider the configuration $F(x) = x^2\Psi(x)$. Its
Stieltjes moments (for symmetric solutions) will satisfy the moment equation 
derived from
Eq.(28) :

$$gw(\rho+2) = -mw(\rho) + Ew(\rho-1) + 2\rho(2\rho-1)w(\rho-2),\eqno(34)$$
\smallskip\noindent $\rho \geq 2$, where $w(\rho) \equiv u(\rho+1)$.
This corresponds to an effective $1+m_s = 4$ order relation since  the
missing moments $\{w(0),w(1),w(2),w(3)\}$ must be specified before all the other
moments
can be generated.

\smallskip\indent However, application of EMM, to the above moment equation,
 will not generate any discrete state energy bounds.
The principal reason for this is that  the same moment equation ensues
if we multiply both sides of (the modified Schrodinger equation)
$$ \Big( -\partial_x^2 + mx^2 + gx^6 - E \Big )\Psi(x) = {\cal D}(x),\eqno(35)
$$
\smallskip\noindent by $x^{p+4}$, $p \geq 0$, provided
 ${\cal D}(x)$ is a (symmetric)
distribution which is projected out when multiplied by
$x^4$. Thus, we can have ${\cal D}(x) = A\delta(x) + B\delta''(x)$, where $A$ 
and $B$ are arbitrary.
It is reasonable to expect that Eq.(35) admits many bounded, positive,
 solutions,
 for
arbitrary $E$; thereby explaining the lack of any EMM generated bounds for the 
$w$
moments.

\smallskip\indent In general, when generating a
moment equation, we are free to multiply both sides of the the Schrodinger
equation by expressions of the form $x^pT(x)$ (where $p \geq 0$)
so long
 as all the zeroes of $T$ 
are zeroes of the desired physical solution (i.e. if $T(x_z) = 0$, then
$\Psi(x_z) = 0$).  If
this is not satisfied, then the resulting ME relation will fail to distinguish
between the true Schrodinger equation, and that modified by additional distribution terms supported at zeroes of $T$. 

In accordance with the above, whereas $T(x) = x^4$ generates a moment equation 
that yields no discrete 
states,
 the function $T(x) = 1+x^2$ does generate the ground state solution. Applying
$x^{2\rho}(1+x^2)$ to both sides of Eq.(9),
we obtain the Stieltjes moment equation
\smallskip\noindent
$gu(\rho+4) = -gu(\rho+3) - mu(\rho+2) + (E-m) u(\rho+1)$
$$ + [E+2(\rho+1)(2\rho+1)]u(\rho)
+2\rho(2\rho-1)u(\rho-1).\eqno(36)$$
\smallskip\noindent This is a $1+m_s = 4$ order relation. Application of EMM 
generates
the ground state energy (although at a slower convergence rate):
$1.4356178 < E_{gr}  < 1.4356185$, utilizing Stieltjes moments  $\{u(\leq 30)\}$
.

A more instructive example is that of the first excited state for the sextic anharmonic
oscillator. The wavefunction will be
 of the form $\Psi_{exc}(x) = x\Upsilon_1(x)$, where
$\Upsilon_1(x) > 0$, and $\Upsilon_1(-x) = \Upsilon_1(x)$, for $x \in \Re_x$. We can transform the Schrodinger equation into an equation for $\Upsilon_1(x)$:
$$ -\epsilon\Big( {2\over x}\Upsilon_1'(x) + \Upsilon_1''(x) \Big ) + [mx^2+gx^6]\Upsilon_1(x) = E\Upsilon_1(x).\eqno(37)$$
Integrating both sides with respect to $x^{2\rho}$ will yield the corresponding
Stieltjes moment equation; however, it will involve (for $\rho = 0$)
 the non-($\Upsilon_1$)moment expression
$\int_{-\infty}^{+\infty}dx {{\Upsilon_1'(x)}\over x},$ which is finite. Although a corresponding EMM analysis can be implemented, it will require a modification
of the conventional EMM formalism, as previously defined.

An alternate approach is to simply take $T(x) \equiv x$, and
 work with the configuration $\Xi(x) \equiv x\Psi(x) = x^2\Upsilon_1(x)$. 
The Stieltjes-$\Xi$ moments are $w(\rho) \equiv \int_{-\infty}^{+\infty}
dx \ x^{2\rho}\Xi(x)$, for $\rho \geq 0$. In terms of the Hamburger moments,
these become $w(\rho) = \mu(2\rho+1)$. 
If we return to the Hamburger ME relation in Eq.(11),
and take $p = 2\rho+1$,
we obtain the desired $w$-Stieltjes equation:
$$-\epsilon 2\rho(2\rho+1) w(\rho-1) + m w(\rho+1) + g w(\rho+3) = E w(\rho),\eqno(38)$$
$\rho \geq 0$. For the case $\epsilon = m = g = 1$, working with the first 30
Stieltjes moments, we obtain the bound
$$ 5.033395937697 < E_1 < 5.033395937709.\eqno(39)$$

For the ground state wavefunction, the function $T(x)$ cannot be zero except
where $\Psi_{gr}(x)$ is zero. For problems  defined on a compact domain,
this means that $T$ can be zero only at the boundary, where the ground state
wavefunction will, generally, be zero. We discuss this in the following
section.

\vfil\break
\bigskip\centerline {\bf IV. A Hausdorff Moment Problem: The Infinite Square Well }

\smallskip\indent We now consider the infinite square well problem

$$-\partial_x^2 \Psi(x) = E \Psi(x),\eqno(40)$$
\smallskip\noindent where $\Psi(\pm L) = 0$. The Hamburger
moments are $\mu(p) = \int_{-L}^{+L}dx \ x^p\Psi(x)$. 
For symmetric configurations,  we have $\mu(2\rho) \equiv u(\rho) = 
\int_0^{L^2}dy \ y^\rho \Phi(y)$, where $y \equiv x^2$ and
$\Phi(y) \equiv {{\Psi(\sqrt{y})}\over {{\sqrt{y}}}}$. In terms
of these {\it Hausdorff} moments, the corresponding moment
equation becomes

$$  -2\rho(2\rho-1) u(\rho-1) - 2L^{2\rho}\Psi'(L) = E\ u(\rho),\eqno(41)$$
\smallskip\noindent $\rho \geq 0$. It involves the boundary terms at 
$\pm L$. 

Relative to the Stieltjes problem, the Hausdorff moment problem introduces
more constraints to the previous Stieltjes (HH) inequalities. We can,
 intuitively,
derive these by assuming that $\Phi(y)$ is nonnegative on $[0,\infty)$. This 
is what the Stieltjes (HH) constraints in Eq.(31) guarantee (i.e. if 
we pretend that the Hausdorff moments are actually Stieltjes moments).

In order to further constrain such a function  so that it be zero on
the interval $[L^2,\infty)$, we must require that $(L^2-y)\Phi(y)$ be nonnegative on $[0,\infty)$:

$$ \int dy \ (L^2 - y) \Big ( \sum_{i = 0}^IC_i y^i\Big)^2 \Phi(y) > 0, \eqno(42)$$

\smallskip\noindent $I < \infty$. That is:
$$ \sum_{i_1,i_2 = 0}^{I} C_{i_1}\Big(L^2u(i_1+i_2)-u(1+i_1+i_2)\Big)             C_{i_2} > 0.\eqno(43)$$
This is the third set of HH constraints that
must be added to those in Eq.(31), for the Hausdorff problem. 

We can summarize all the Hausdorff-HH relations by

$$ \sum_{i_1,i_2 = 0}^{I} C_{i_1}^{(\sigma)}\Big(\Gamma_{\sigma}^{(1)}u(i_1+i_2)+ 
\Gamma_{\sigma}^{(2)}u(1+i_1+i_2)\Big)
  C_{i_2}^{(\sigma)} > 0,\eqno(44)$$
for $\sigma = 0,1,2$,
where
$$\Gamma_{\sigma}^{(1)} = \cases{ 1,\ \sigma = 0 \cr
	                          0,\ \sigma = 1 \cr
			          L^2,\ \sigma = 2 \cr},\eqno(45)$$
and
$$\Gamma_{\sigma}^{(2)} = \cases{ 0,\ \sigma = 0 \cr
                                  1,\ \sigma = 1 \cr
                                  -1,\ \sigma = 2 \cr},\eqno(46)$$
for all nontrivial $C^{(\sigma)}$'s, and $I \geq 0$.

We outline how the above constraints lead to the quantization of the ground 
state.
\smallskip\indent Let $L = 1$, and $A \equiv 2\Psi'(L)$. Then
$u(0) = -{A\over E} > 0$, and all the remaining moments can be generated, once $
A$ is
normalized.
The first three Hausdorff-HH conditions ($I = 0$) become:

$$ {\it Hausdorff-HH\ relations}(I = 0)\ \rightarrow \cases {  u(0) > 0; \cr
         u(1) > 0; \cr
	u(0) - u(1) > 0 \cr}.
\eqno(47)$$
Combining $Eu(1) = -2u(0)-A$, and $u(1) > 0$,  yields
$-2{{u(0)}\over E} + u(0) > 0$, or  ${2\over E} < 1$. The third  inequality,
$ {{u(1)}\over {u(0)}} < 1$, yields $ -{2\over E} + 1 < 1$, or $E > 0$; hence
the lower bound 
$$ 2 < E.\eqno(48)$$
Having established the positivity of $E$, we are free
to
impose the normalization $A = -1$, hence $u(0) = {1\over E}$. Thus, the 
ME relation effectively becomes a zero missing moment problem, with
$m_s = 0$.  We can procede with a numerical determination of the ground state energy.

\smallskip\indent For problems corresponding to $m_s = 0$, we do not have
to implement the linear programming based, EMM, formulation. Instead,
 we can work with the
nonlinear
HH inequalities (which are the relations usually cited in the literature)
corresponding
to the quadratic form relations given previously. That is, the Hausdorff-HH 
linear (in the moments) constraints, are equivalent to the nonlinear (in the
moments) determinantal relations:

$$ Det\Big ( \Delta^{(I)}_\sigma \Big ) > 0, \eqno(49)$$
\smallskip\noindent where the various HH matrices are

$$\Delta^{(I)}_{\sigma;i_1,i_2} = 
\Gamma_{\sigma}^{(1)}u(i_1+i_2)
+
\Gamma_{\sigma}^{(2)}u(1+i_1+i_2),\eqno(50)$$
for $\sigma = 0,1,2$,
and $0 \leq i_1,i_2 \leq I$.
\smallskip\indent The numerical evaluation of these  inequalities  yields the
 bounds
$$ 2.4674010541 < E <   2.4674011008,\eqno(51)$$
utilizing all the HH determinants corresponding to the first seven moments:
$\{u(\leq 6)\}$. This compares exceptionally well (up to seven decimal places)
 with the true answer, $E = ({\pi\over 2})^2$.

\bigskip\noindent{\underbar{\it  Moment Equations with No Boundary Terms}}
\smallskip\indent  In practice, particularly for multidimensional applications,
 we prefer to work with moment equations that do not involve any boundary terms.
For the infinite square well case, we can do so by multiplying both 
sides of the corresponding Schrodinger equation
by $x^{p}T(x)$, where $T(\pm 1) = 0$.
The ensuing moment equation 
will not involve any boundary terms because the kinetic
energy term  becomes

$$\int_{-1}^{+1} dx G(x) \Psi''(x) =  G\Psi'|_{-1}^{+1} - G'\Psi|_{-1}^{+1} +
\int_{-1}^{+1} dx \ G''(x)\Psi(x),\eqno(52)$$
\smallskip\noindent where $G(x) \equiv x^pT(x)$.
Since both $G$ and $\Psi$ are zero at the boundary, no boundary terms 
will contribute to the ensuing ME relation.

\smallskip\indent As an example, let $T(x) = 1-x^2$, for the 
$L = 1$ case. Applying $x^{2\rho}(1-x^2)$
 to
both sides of the infinite square well problem yields the Hausdorff moment 
equation

$$ Eu(\rho+1) = [E-(2\rho+2)(2\rho+1)]u(\rho) + 2\rho(2\rho-1) u(\rho-1), \eqno(
53)$$
\smallskip\noindent for $\rho \geq 0$. Application of EMM duplicates the 
bounds previously cited.

\vfil\break
\centerline {\bf V. The Infinite Quantum Lens Problem}
The quantum lens geometry, as shown in Fig. 1,  is bounded by the $z = R_1$ plane, and the sphere
of radius $R_2$:
$$ {\it Lens \ Domain} = \{ z \geq R_1 \} \bigcap \{ r \leq R_2 \},\eqno(54)$$
where $R_1 < R_2$.

In a cylindrical coordinate representation,  the Schrodinger equation for
the infinite quantum lens potential problem
becomes (in energy units $E_0 = {{\hbar^2}\over {2ma^2}}$, and length 
in units of the radius,  $a = \sqrt{R_2^2-R_1^2}$ ):

$$-\Big( {1\over \rho}\partial_\rho(\rho\partial_\rho\Psi) + {1\over{\rho^2}}\partial_\phi^2
\Psi + \partial_z^2\Psi \Big ) = E \Psi(\rho,\phi,z),\eqno(55)$$
\smallskip\noindent where ($r^2 = \rho^2+z^2$, note that we will be working with  the $r^2$ and $z^2$ coordinates)

$$R_1^2\leq r^2 \leq R_2^2, \ {\rm and} \ 
R_1^2 \leq z^2 \leq r^2,\eqno(56)$$
for $ z > 0$.
The boundary condition on the wavefunction is 
$$ \Psi(\rho,\phi,z) = \cases { 0,\ z^2 = R_1^2 \cr
				0,\ r^2 = R_2^2 \cr}.\eqno(57)$$ 
\smallskip\indent The radii $R_1$ and $R_2$ can be redefined in terms of the 
quantum lens
parameters $a$ and 
$b$,  where
$$ a^2 = R_2^2 - R_1^2, \ {\rm and} \ 
 b = R_2-R_1,\eqno(58)$$
\smallskip\noindent or, alternatively,
$$ R_2 ={{ {{a^2}}+ b^2}\over {2b}},  \ {\rm and} \ R_1 ={{ {{a^2}}- b^2}\over {2b}}.\eqno(59)$$
\smallskip\indent The lens domain transforms into a triangular domain in  the 
$\{r^2,z^2\}$ coordinate space, or, equivalently,

$$ \omega \equiv R_2^2 - r^2, \ {\rm and} \
 \nu = z^2 - R_1^2. \eqno(60)$$
\smallskip\noindent The corresponding domain is
$$ 0 \leq \omega \leq a^2 \ {\rm and} \ 0 \leq \nu \leq a^2-\omega.\eqno(61)$$

Equation (55) is axially symmetric, and the solutions assume the form
$\Psi(\rho,\phi
,z) =
e^{-im\phi}\psi(\rho,z)$.

\smallskip\indent In the $\{\omega,\nu\}$ coordinate system,  
 the Schrodinger equation becomes (i.e. first transform 
into $\{\rho^2,z^2\}$ coordinates, then into $\{r^2,z^2\}$, and finally into
 $\{\omega,\nu\}$):
\smallskip\noindent
$-4\Big( (R_2^2-\omega)\partial_\omega^2-{3\over 2}\partial_\omega 
-2(R_1^2+\nu)\partial_\omega\partial_\nu + {1\over 2} \partial_\nu
+ (R_1^2 + \nu)\partial_\nu^2 \Big )\psi(\omega,\nu)$
$$ + {{m^2}\over {a^2-\omega-\nu}}\psi(\omega,\nu)
= E \psi(\omega,\nu).\eqno(62)$$
\smallskip\indent The boundaries $r^2 = R_2^2$ and $z^2 = R_1^2$ become
$\omega = 0$ and $\nu = 0$. According to Eq.(61), the $\{\omega,\nu\}$ physical domain is restricted
to the lower left triangle of the $[0,a^2]\times[0,a^2]$ square region. The hypotenuse of 
this triangle corresponds to $a^2-\omega -\nu = \rho^2 = 0$. The wavefunction is not 
zero along it; although it is zero along  $\omega = 0$ and $\nu = 0$.

\smallskip\indent Although we shall work within the $\{\omega,\nu\}$ coordinates, 
in order to derive the necessary moment equations, we note that we can rewrite the
above equation in terms of  the coordinates $\xi \equiv \omega+\nu$ and 
$\eta = \omega - \nu$. The derivatives become
 $\partial_\omega = \partial_\xi + \partial_\eta$,
 $\partial_\nu    = \partial_\xi - \partial_\eta$. The $\rho = 0$ boundary corresponds to 
$\xi = a^2$. In terms of these new coordinates, the Schrodinger equation becomes
\smallskip\noindent
$-4\Big( [a^2-\xi]\partial_\xi^2 + 2[a^2-\xi]\partial_\xi\partial_\eta 
+ [R_2^2+3R_1^2+(\xi-2\eta)]\partial_\eta^2 -\partial_\xi-2\partial_\eta \Big)\psi$
$$+{{m^2}\over {a^2-\xi}}\psi = E \psi.\eqno(63)$$
\smallskip\noindent We shall refer to the various function coefficients of the derivative 
operators in Eq.(63) (i.e. $\sum_{i,j}C_{i,j}\partial_\xi^i\partial_\eta^j$) by 
$$C_{i,j}(\xi,\eta) = \cases { a^2-\xi, i = 2, j = 0 \cr
				2[a^2-\xi], i = 1, j = 1 \cr
				R_2^2+3R_1^2+(\xi-2\eta) , i = 0, j = 2 \cr
				-1, i = 1, j = 0 \cr
				-2, i = 0, j = 1 \cr }. \eqno(64)$$

\smallskip\indent Our objective is to derive, for a given quantum number 
$m$, a moment equation for  Eq.(62), involving the moments
$$u(p,q) \equiv \int_0^{a^2}d\omega \int_0^{a^2-\omega} d\nu \ \omega^p\nu^q\psi(\omega,\nu),
\eqno(65)$$
and no boundary terms (in a manner consistent with the requirements defined 
in Sec. III).

\smallskip\indent In order to achieve the above, for the $m = 0$ case, we will have to 
multiply both sides of Eq.(62) by 
$G(\omega,\nu) = \omega^p\nu^q$, where $p,q \geq 1$. We note that $G(\omega,\nu) = 0$, along both boundaries
$\omega = 0$ and $\nu = 0$, where $\psi = 0$.  
Integrating over the triangular domain in $\{\omega,\nu\}$ does not introduce 
any boundary terms at all, not even along the $\rho = 0$ boundary, where $\psi \neq 0$. We prove this, below, for each of the contributing terms in the ME
relation.

\smallskip\noindent {\it (i)} The terms 
$G(\omega,\nu) C_{0,2}(\xi,\eta) \partial_\eta^2\psi$ and 
$G(\omega,\nu) C_{0,1}(\xi,\eta) \partial_\eta\psi$, do not introduce any boundary terms
since those generated by integration by parts (in the $\eta$ direction) 
correspond to points where  $G(\omega,\nu) = 0$ and $\psi = 0$.

\smallskip\noindent {\it (ii)} The integration by parts of $G(\omega,\nu)C_{1,1}(\xi,\eta)\partial_\xi
\partial_\eta\psi$ reduces to 
$$\partial_\xi(GC_{1,1}\partial_\eta\psi) - 
\partial_\eta(\psi \partial_\xi(GC_{1,1})) + \psi \partial_\xi\partial_\eta(GC_{1,1}).$$
The boundary terms produced by the first term (along the $\xi$ direction) are zero since 
at one point (corresponding to either $\omega = 0$ or 
$\nu = 0$) we have $G(\omega,\nu) = 0$, while at the other (corresponding to $\xi = a^2$), 
we have $C_{1,1} = 0$. The boundary terms from the second term  are also zero, since at
both ends (along the $\eta$ direction) we have $\psi = 0$.

\smallskip\noindent {\it (iii)} The integration by parts for 
$G(\omega,\nu)[C_{2,0}(\xi,\eta)\partial_\xi^2\psi - \partial_\xi\psi]$  gives us
\smallskip\noindent
$$\partial_\xi(GC_{2,0}\partial_\xi\psi)-\partial_\xi(\psi\partial_\xi(GC_{2,0}))  
-\partial_\xi(G\psi) + [\partial_\xi^2(GC_{2,0}) + \partial_\xi G]\psi.$$
The first term introduces no boundary terms (along the $\xi$ direction) because at
 one endpoint we have $G(\omega,\nu) = 0$, while at the other $C_{2,0} = 0$. The second and third terms
have no boundary term at the point corresponding to $\psi = 0$. However, at 
$\xi = a^2$, since $\partial_\xi C_{2,0} = -1$, we obtain a cancellation
 between 
the only surviving boundary terms. This concludes the proof that no boundary terms 
arise for the $m = 0$ case.

\smallskip\indent For the $m \neq 0$ case,  we take $G(\omega,\nu) = \omega^p\nu^q(a^2-\omega-\nu)$,
 for $p,q \geq 1$.
The preceding argument still holds, although 
the final cancellation of both terms is unnecessary because
of the additional $(a^2-\xi)$ factor introduced through the modified $G(\omega,\nu)$.

\smallskip\indent Given the above, it is now straightforward to generate the  
required
moment equation.
\vfil\break
\bigskip\noindent \underbar{\it The $m = 0$ Moment Equation}
\bigskip\indent  The moment equation for the $m = 0$ case is
\smallskip\noindent
$-{E\over 4}u(p,q) = R_2^2\ p(p-1)\ u(p-2,q) - [p^2 + {{3p}\over 2} + 2pq]\ u(p-1,q)$
$$-2R_1^2\ p q\ u(p-1,q-1) +[q^2+ {q\over 2}]\ u(p,q-1) + R_1^2q(q-1)\ u(p,q-2),\eqno(66)$$
\smallskip\noindent for $p,q \geq 1$. 
\smallskip\indent The {\it missing moments}
$\{u(0,0),\dots,u(N,0)\}$ and $\{u(0,1),\ldots,u(0,N)\}$, generate all the moments
within the square grid $[0,N]\times [0,N]$. We can index the missing moments according to
$\chi_0 \equiv u(0,0)$, $\chi_1 \equiv u(1,0)$, $\ldots$, $\chi_N \equiv u(N,0)$, $\chi_{N+1} 
\equiv u(0,1)$, $\ldots$, $\chi_{2N} \equiv u(0,N)$. 
\smallskip\indent We can then determine the 
energy dependent coefficients linking the moments to the missing moments
$$u(p,q) = \sum_{\ell = 0}^{m_s = 2N} M_E(p,q,\ell) \chi_\ell.\eqno (67)$$
The $M_E$ coefficients satisfy the moment equation with respect to the 
$p,q$ indices. In addition, $M_E(p_{\ell_1},q_{\ell_1},\ell_2) = \delta_{\ell_1,\ell_2}$, where $(p_{\ell_1},q_{\ell_1})$ denotes the coordinates of the 
missing moments.
\smallskip\indent As explained in the previous examples,
 one can impose a normalization condition
of the form $\sum_{\ell=0}^{2N}\chi_\ell = 1$, constraining $\chi_0$. Incorporating this
within the above relation we have  

$$u(p,q) = {\hat M}_E(p,q,0) + \sum_{\ell = 1}^{m_s = 2N} {\hat M}_E(p,q,\ell) \chi_\ell,\eqno (68)$$
where 
$${\hat M}_E(p,q,\ell) = \cases{ M_E(p,q,0),\ \ell = 0 \cr
M_E(p,q,\ell) -  M_E(p,q,0), \ \ell \geq 1 \cr}.\eqno(69)$$

\smallskip\indent From the positivity theorems of the Moment Problem, we have
to impose the moment constraints arising from the integral relations

$$\int\int d\omega d\nu\ \Omega_\sigma(\omega,\nu)\ 
\Big ( \sum_{i,j\in [0,I]^2} {\tilde C}_{i,j} \omega^i\nu^j \Big )^2 \psi
(\omega,\nu) > 0, \eqno(70)$$
\smallskip\noindent for arbitrary ${\tilde C}$'s (not all zero), where

$$ \Omega_\sigma(\omega,\nu) = \cases { 1,\ \sigma = 0 \cr
					\omega,\ \sigma = 1 \cr
					\nu,\ \sigma = 2, \cr
					a^2-\omega-\nu,\ \sigma = 3 \cr }. \eqno(71)$$
\smallskip\noindent It is implicitly assumed that $\psi$ is zero outside the triangular 
domain of interest. 
\smallskip\indent These integral inequalities become linear inequalities, with respect
to the $u$-moments:

$$ \sum_{i_1,j_1}\sum_{i_2,j_2} {\tilde C}_{i_1,j_1} 
\Big(\sum_{n=1}^3f_{\sigma,n}u(\lambda_{1;\sigma,n} + 
i_1+i_2,\lambda_{2;\sigma,n} + j_1+j_2) \Big ) {\tilde C}_{i_2,j_2} > 0, \eqno(72)$$
\smallskip\noindent where
$$ f_{\sigma,n} = \cases {1,0,0, \ for \ \sigma = 0 \cr
	                  0,1,0, \ for \ \sigma = 1 \cr
			  0,0,1, \ for \ \sigma = 2 \cr
			  a^2,-1,-1 \ for \ \sigma = 3 \cr }, \eqno(73)$$
\smallskip\noindent and  the  $\lambda$'s associated with nonzero $f_{\sigma,n}$'s
$$ (\lambda_{1;\sigma,n},\lambda_{2;\sigma,n}) = \cases { 
				(0,0), \ for \ \sigma =\ 0, n = 1 \cr
				(1,0), \ for \ \sigma =\ 1, n = 2 \cr
				(0,1), \ for \ \sigma =\ 2, n = 3 \cr
				(0,0), (1,0), (0,1), \ for \ \sigma = 3 \cr }.\eqno(74)$$

\smallskip\indent If one defines a coordinate pair sequence $(i_l,j_l) \in [0,I]\times [0,I]$,
then the set of points covered by $(i_{l_1},j_{l_1}) + (i_{l_2},j_{l_2}) + 
(\lambda_{1;\sigma,n},\lambda_{2;\sigma,n})$, lie within a square grid $[0,N]^2$, where
$N = 2I+1$. All the moments within this grid will be generated by the missing moments, 
previously defined.
\smallskip\indent One procedes by substituting the moment - 
missing moment relation in Eq.(68) into Eq.(72).
 This defines an infinite set of linear inequalities
in the missing moments, and can be analyzed through the
linear programming based
EMM algorithm.  The EMM numerical analysis generates a finite number of 
 optimal ${\tilde C}$'s which  
determine if, to order $I$, the (normalized) 
inequalities in Eq.(72) have a solution set, 
${\cal U}_{E}^{(I)}$, for the specified $E$ value. The feasible energies define the 
converging lower and upper bounds. 

\smallskip\indent In Tables I and II we give some results of our 
approach, as a function of the ratio ${b\over a}$, for the $m = 0$ case. Already, for $I = 2$ ($m_s = 10$), and 
$I = 3$ ($m_s = 14$), we obtain very good bounds for the ground state energy, even for 
small lens thickness.
As the ratio ${b\over a}$ 
becomes smaller, the lens becomes thiner, with maximum thickness $b$, and base  diameter
$2a$.

In Fig. 2, we compare the normalized ground state energy ($E\over E_0$) , for 
$m = 0$, as a function of the ratio $b\over a$, obtained by three methods:
\smallskip\indent ({\it i}) exact numerical solution (solid line) of Eq.(62);
\smallskip\indent ({\it ii}) perturbation theory (dashed lines), based on a conformal transformation into a semi-spherical shape (Rodriguez et al (2001));
\smallskip\indent ({\it iii}) EMM analysis, as reported in Table I (
solid black dots for the lower and upper
bounds, when the ``bounding" bars become too small).

It can be seen that very good agreement is obtained between the exact solution
and the EMM bounds, for ${b\over a} < 0.7$. The perturbation results yield 
better agreement with the exact, numerical solution, for ${b\over a} > 0.8$.

\vfil\break

$$ \table{\bf Table I:  Ground state energy bounds ($m = 0, I = 2$)
} 
${b\over a}$       !Bounds       !${b\over a}$       !Bounds \rr
.80           ! 6  $< \ E \ < $ 30 ! .20            !292.6 $< \ E \ < $  295.5           \rr
.75             !26 $< \ E \ < $ 33 !.15               !496 $< \ E \ < 506$            \rr
.70              !33.8  $< \ E \ < $ 35.2 !.10               ! 1064 $< \ E \ < 1093 $         \rr
.65               !38.8  $< \ E \ < $39.4  ! .09             ! 1300 $< \ E \ < 1340$ \rr
.60                !44.33 $< \ E \ < $ 44.42   !.08          ! 1630 $< \ E \ < 1680 $        \rr
.55                !50.98$< \ E \ < $       51.03 !.07       ! 2110 $< \ E \ < 2180$        \rr
.50              !59.54 $< \ E \ < $ 59.58 ! .06            ! 2850 $< \ E \ < 2950  $\rr
.45                !70.85 $< \ E \ < 70.92 $ !.05             ! 4070 $< \ E \ < 4210 $   \rr
.40               ! 86.32$< \ E \ < $ 86.44 ! .04              ! 6320 $< \ E \ < 6520 $\rr 
.35             !  108.30$< \ E \ < $  108.70 !.03              !11160 $< \ E \ < 11520$       \rr
.30                !141.60  $< \ E \ < $142.20 !.02              !24950 $< \ E \ < 25800$ \rr
.25                !195.60  $< \ E \ < $196.80 ! .01              !99170 $< \ E \ < 102970$
\caption{\sl  }$$

$$\table{\bf Table II :  Ground state energy bounds ($m = 0, I = 3$)}
${b\over a}$       !Bounds ! ${b\over a}$       !Bounds     \rr
.40  !      86.37 $ < E < $ 86.39 ! .10  !    1077 $ < E < $1080 \rr
.30  !     141.90 $ < E < $ 141.94! .05 !   4120 $ < E <$ 4135 \rr
.20  !     293.92 $ < E < $ 294.00 !     !                  
\caption{\sl  }$$

\vfil\break
\centerline{\bf Conclusion}

We have presented a preliminary analysis of the infinite quantum lens 
potential problem in terms of EMM theory. It is anticipated that
a different choice of coordinates will improve the bounds for
${b\over a} > O(.7)$. These considerations, and the extension of
the EMM bounding method to the lowest energy state, 
within each $m \neq 0$ class, are the subject of 
ongoing research, and will be communicated in a future work.

\vfil\break
\centerline {\bf Acknowledgements}
\bigskip\indent This work was supported by a grant (HRD-9632844) from the National Science Foundation 
through the Center for Theoretical Studies of Physical Systems. 
\vfil\break
\centerline {\bf References}
\bigskip\noindent
Akhiezer N I 1965 {\it The Classical Moment Problem and Some
Related Questions in Analysis} (Oliver and Boyd, Edinburgh).
\bigskip\noindent Bender C M and Orszag S A 1978 {\it Advanced 
Mathematical Methods for Scientists and Engineers} (New York: 
McGraw-Hill).
\bigskip\noindent
Chvatal V 1983 {\it Linear Programming} (Freeman, New York).
\bigskip\noindent
Handy C R and Bessis D 1985 Phys. Rev. Lett. {\bf 55} 931.
\bigskip\noindent
Handy C R,  Bessis D,  and  Morley T D 1988a Phys. Rev. A {\bf 37} 4557.
\bigskip\noindent
Handy C R,  Bessis D, Sigismondi G, and Morley T D 1988b Phys. Rev. Lett.
{\bf 60} 253.
\bigskip\noindent
Drexler H, Leonard D, Hansen W, Kotthaus J P, and Petroff P M 1994 
Phys. Rev. Lett. {\bf 73} 2252.
\bigskip\noindent
Fafard S, Leon R, Leonard D, Merz J L, and Petroff P M 1994 Phys. Rev. B 
{\bf 50} 8086.
\bigskip\noindent LeGuillou J C and Zinn-Justin J 1983 Ann. Phys. (N.Y.)
{\bf 147} 57.
\bigskip\noindent
Lee S, Kim J C, Rho H, Kim C S, Smith L M, Jackson H E, Furdina J K, and
Dobrowdska M 2000 Phys. Rev. B {\bf 61} 2405.
\bigskip\noindent
Leonard D, Krishnamurthy M, Reaves C M, Denbaars S, and Petroff P M 1993
Appl. Phys. Lett. {\bf 63} 3203.
\bigskip\noindent 
Leonard D, Pond K, Petroff P M 1994 Phys. Rev. B {\bf 50} 11687.
\bigskip\noindent
Medeiros-Ribeiro G, Pikus F G, Petroff P M, and Efros A L 1997 Phys. Rev. B
{\bf 55} 1568.
\bigskip\noindent
Miller B T, Hansen W, Manus S, Lorke A, and Kotthaus J P 1997 Phys. Rev. B {\bf 56}
6764.
\bigskip\noindent Handy C R and Murenzi R 1998 J. Phys. A {\bf 31} 9897.
\bigskip\noindent
Reed M and  Simon B 1978 {\it Methods of Modern Mathematical
Physics} (Academic, New York), p. 206, Theorem XIII-46.
\bigskip\noindent
Rodriguez A H,  Trallero-Giner C,  Ulloa S E, and 
Marin-Antuna J (2001 to be published in Phys. Rev. B)
\bigskip\noindent
Shohat J A and  Tamarkin J D 1963 {\it The Problem of Moments}
(American Mathematical Society, Providence, R.I.).
\bigskip\noindent
\vfil\break
\centerline {Caption for Tables and Figures}
\bigskip\noindent Table I: Ground state energy bounds ($m = 0, I = 2$).
\bigskip\noindent Table II: Ground state energy bounds ($m = 0, I = 3$).
\bigskip\noindent Fig. 1: Quantum lens geometry of height b and circular
coss section of radius a.
\bigskip\noindent Fig. 2: Ground state energy for a quantum lens
as a function of the ratio ${b \over a}$. The energy is given in units of
$E_0 = {{\hbar^2}\over{2ma^2}}$, as calculated by:
a) exact numerical solution (solid line); perturbation theory  
with respect to the $b\over a$ parameter (dashed line, Rodriguez et al (2001));
 and EMM theory, as given
in Table I (solid black dots). The lower and upper energy bounds are represented by
``bounding" bars (which cannot be depicted, for smaller ${b\over a}$ values) .

\end